\documentclass[preprint,prc,showpacs,preprintnumbers,amsmath,amssymb]{revtex4-1}
\usepackage{graphicx, latexsym, amssymb, amsmath, color, multirow, mathrsfs, booktabs, ifpdf}
\usepackage{dcolumn}
\usepackage{bm}
\usepackage{longtable}
\def\nA{nucleon-nucleus\ }
\def\pA{proton-nucleus\ }

\def\phe6{$^6$He+$p$\ }
\def\pnA{$A(p,n)\tilde{A}_{\rm IAS}$\ }
\def\pnP{$^{208}$Pb$(p,n)^{208}$Bi$^*_{\rm IAS}$\ }
\def\An{$n+\tilde{A}$\ }
\def\Ap{$p+A$\ }
\def\Pb{$^{208}$Pb\ }
\def\pPb{$p+^{208}$Pb\ }
\def\nPb{$n+^{208}$Pb\ }
\def\pn{$(p,n)$\ }
\def\pp{$(p,p)$\ }
\begin{document}
\title{Neutron scattering from $^{208}$Pb at 30.4 and 40.0 MeV and isospin dependence
 of the nucleon optical potential}
\author{R.P. DeVito$^1$\footnote{MSU Technologies, 325 East Grand River, Suite 350
East Lansing, MI 48823}}
\author{Dao T. Khoa$^2$\footnote{khoa@vaec.gov.vn}}
\author{Sam M. Austin$^{1,3}$\footnote{austin@nscl.msu.edu}}
\author{U.E.P. Berg$^1$\footnote{Berg Consult, D-73734 Esslingen, Germany}}
\author{Bui Minh Loc$^2$\footnote{University of Pedagogy,
 Ho Chi Minh City, Vietnam}}
\affiliation{$^{1}$National Superconducting Cyclotron Laboratory and\\
Department of Physics and Astronomy\\
Michigan State University, East Lansing, MI 48824-1321. \\
$^2$Institute for Nuclear Science and Technique, VAEC \\
179 Hoang Quoc Viet, Hanoi, Vietnam. \\
$^3$Joint Institute for Nuclear Astrophysics\\
Michigan State University, East Lansing Michigan 48824-1321.}

\begin{abstract}
\begin{description}
\item[Background] Analysis of data involving nuclei far from stability often requires
optical potential (OP) for neutron scattering.  Since neutron data is seldom
available, while  proton scattering data is more abundant, it is useful to have
estimates of the difference of the neutron and proton optical potentials.
This information is contained in the isospin dependence of the nucleon OP.
Here we attempt to provide it for the nucleon-\Pb system.
\item[Purpose] The goal of this paper is to obtain accurate \nPb scattering data,
and use it, together with existing \pPb and \pnP data, to obtain an accurate estimate of
the isospin dependence of the nucleon OP at energies in the 30-60 MeV range.
\item[Method] Cross sections for \nPb scattering were measured at 30.4 and 40.0 MeV,
with a typical relative (normalization) accuracy of 2-4\% (3\%). An angular range
of 15 to 130 degrees was covered using the beam-swinger time of flight system
at Michigan State University. These data were analyzed by a consistent optical
model study of the neutron data and of elastic \pPb scattering at 45 MeV and 54 MeV.
These results were combined with a coupled-channel analysis of the $^{208}$Pb$(p,n)$
reaction at 45 MeV, exciting the 0$^+$ isobaric analog state in $^{208}$Bi.
\item[Results] The new data and analysis give an accurate estimate the isospin impurity
of the nucleon-\Pb OP at 30.4 MeV, caused by the Coulomb correction to the proton OP.
The  corrections to the real proton OP given by the CH89 global systematics
was found to be only few percent, while for the imaginary potential it was over
20\% at the nuclear surface. Based on the analysis of the measured elastic \nPb
data at 40 MeV, a Coulomb correction of similar strength and shape was also
predicted for the \pPb OP at energy around 54 MeV.
\item[Conclusions] Accurate neutron scattering data can be used in combination
with proton scattering data and \pn charge exchange data leading to the IAS to
obtain reliable estimates of the isospin impurity of the nucleon OP.
\end{description}
\end{abstract}
\date{\today}

\pacs{25.40.Dn; 24.10.Ht; 24.10.Eq}
\maketitle

\section{Introduction}
At energies below 100 MeV, the attraction between a neutron and a proton is stronger
than that between two protons or two neutrons. Consequently, the average interaction
of a proton with an $N>Z$ nucleus is stronger than that of a neutron. In other words,
the nuclear interaction between an incident nucleon and a target with non-zero isospin
has an isospin dependent part. For the nuclear part of the \nA optical potential (OP),
the isospin dependent term is, in the Lane form \cite{La62},
\begin{equation}
 U_N=U_0+4U_1\frac{{\bm t}.{\bm T}}{A}, \label{e1}
\end{equation}
where ${\bm t}$ is the isospin of the incident nucleon and ${\bm T}$ is that of
the target $A$.
The second term of Eq.~(\ref{e1}), known as the Lane potential, contributes to
both the elastic ($p,p$) and ($n,n$) scattering as well as to the
charge exchange ($p,n$) reaction \cite{Sat64,Sat83}. A knowledge of $U_1$ is
of fundamental interest for  studies of nuclear phenomena in which neutrons and
protons participate differently (isovector modes). Many previous estimates of $U_1$
(see, for example, Refs.~\cite{Rap79,Pat76,Jeu77}) involved a comparison of the
nucleon OP from a range of nuclei with different values of the asymmetry parameter
$\varepsilon=(N-Z)/A$ or the Distorted Wave Born Approximation (DWBA) analysis
of $(p,n)$ reactions exciting the isobaric analog state (IAS). However, these
approaches are subject to serious uncertainties. For example, in the comparison
of elastic nucleon scattering from different nuclei one must make assumptions
\cite{Rap79} about the variation of nuclear geometry with $A$ and $\varepsilon$.
Moreover, the contribution of the Lane potential $U_1$ to the elastic nucleon
scattering cross section is relatively small \cite{Kho02,Kho03}. In contrast,
$U_1$ entirely determines the (Fermi-type) $\Delta J^\pi=0^+$ transition strength
of the \pn reaction exciting the IAS; this reaction is, therefore, a sensitive
probe of the isospin dependence of the \pA OP \cite{Kho07}. However, for
\pn reactions, a change in Re$~U_1$ can be approximately compensated by a change
in Im$~U_1$ \cite{Kunz}, and the determination of $U_1$ remains ambiguous
without additional information or constraints.

It is in principle possible to avoid these uncertainties by extracting $~U_1$ from a
consistent study of the elastic proton and neutron scattering and the charge
exchange $(p,n)$ reaction on the same target nucleus, at the same energy. We recall
here briefly the consistent isospin coupling scheme  \cite{Sat83} for the elastic
\nA scattering and charge exchange \pn reaction exciting IAS. For the isospin
projections $T_z=(N-Z)/2$ of the target nucleus $A$ and $\tilde{T_z}=T_z-1$
of the \emph{isobaric analog nucleus} $\tilde{A}$, and denoting the neutron-
and proton scattering states by $|nA\rangle$ and $|pA\rangle$, respectively,
the neutron and proton optical potentials are given by the diagonal matrix
elements of potential (\ref{e1})
\begin{eqnarray}
 U_p=\langle pA|U_N|pA\rangle&=&U_0-\frac{2}{A}T_zU_1
 =U_0-\varepsilon U_1,  \label{e3a} \\
 U_n=\langle nA|U_N|nA\rangle&=&U_0+\frac{2}{A}T_zU_1
 =U_0+\varepsilon U_1,  \label{e3b}
\end{eqnarray}
Similarly, the transition matrix element or \pn form factor (FF) for the charge
exchange \pnA reaction exciting IAS is
\begin{equation}
 F_{pn}=\langle n\tilde{A}|4U_1\frac{{\bm t}.{\bm T}}{A}|pA\rangle
 =\frac{2}{A}\sqrt{2T_z}U_1=2\sqrt{\frac{\varepsilon}{A}}U_1.
 \label{e4}
\end{equation}
If the neutron and proton optical potentials at a given energy are well
determined from the optical model (OM) analysis of the corresponding
elastic data, then the isovector term of the nucleon OP can be obtained
directly from Eqs.~(\ref{e3a})-(\ref{e3b}) as
\begin{equation}
U_1=\frac{(U_n-U_p)}{2\varepsilon}.
 \label{e5}
\end{equation}

Unfortunately, isospin is not a good quantum number in the
Coulomb field of the nucleus since this field slows the incident proton and affects
the strength and shape of the \pA OP. It is necessary, therefore, to add Coulomb
corrections $\Delta E_C$ to the incident proton energy and
$\Delta U_C$ to $U_p$ to separate the main effects of the Coulomb field
so that isospin is a good quantum number for the remainder of the nucleon OP, namely,
\begin{eqnarray}
 U_p &=& U_0 - \varepsilon U_1 + \Delta U_C,  \label{e6a} \\
 U_n &=& U_0 + \varepsilon U_1.   \label{e6b}
\end{eqnarray}
Then the Coulomb correction term must be determined from
\begin{equation}
 \Delta U_C=U_p-U_n + 2\varepsilon U_1 \Longleftrightarrow
 U_1-\frac{\Delta U_C}{2\varepsilon}=\frac{U_n-U_p}{2\varepsilon}. \label{e7}
\end{equation}
In general, Eq.~(\ref{e7}) has 4 unknowns (the real and imaginary parts of
$\Delta U_C$ and $U_1$), that, as we shall see, cannot be determined unambiguously
even if the \emph{complex} neutron and proton optical potentials are well
determined.

There are ``global" systematics of the OP parameters deduced from the extensive
OM analyses of nucleon elastic scattering, for example, those by Becchetti and
Greenlees \cite{BG69}, by Koning and Delaroche \cite{Kon03}, and by
Varner {\it et al.} \cite{Va91}. In the work described here, we rely on the CH89
global model by Varner {\it et al.}. The CH89 optical potentials cover a
wide range of energies and target masses, and are parametrized using Woods-Saxon (WS)
forms.  The resulting systematics are often used to predict
the nucleon OP when elastic scattering data are not available or cannot be
measured, as is the case for many unstable nuclei near the dripline. Given the
large isospins of dripline nuclei, it is important to estimate accurately the
isospin dependence of the nucleon OP (or equivalently, the Coulomb corrections
to that OP) before applying it in studies of nuclear reactions or of
astrophysical phenomena. So far, the empirical
isospin dependence of the nucleon OP has been deduced \cite{BG69,Va91,Kon03}
based mainly on the OM analyses of the proton and neutron elastic scattering,
adopting some simple treatment of the Coulomb correction terms
$\Delta E_C$ and $\Delta U_C$.

Before the present measurement of neutron scattering on  $^{208}$Pb, detailed elastic
\nPb scattering data were available only at energies up to 26 MeV. This energy range,
 however, does not overlap that of much of the precise proton data and furthermore
is too small to establish clearly the energy dependence of the neutron OP.
The measurements at 30.4 and 40.0 MeV described here have greatly expanded the energy
range for the neutron data and provided the most accurate and detailed data for
neutron scattering above 26 MeV from any $N \neq Z$ nucleus. After our measurement,
several experiments on the elastic \nPb scattering have been carried out and
the neutron scattering data were measured at 65 MeV \cite{Hjo94}, in
the forward angular region but over a wide range of the neutron incident energy
from 65 to 225 MeV \cite{Osb04}, and at the neutron energy of 96 MeV \cite{Klu03}.
Together with the present data, one has now a good database of the elastic \nPb
scattering data to study the energy dependence of the neutron OP. In the present
paper we show that the accurate neutron scattering data can also be used in
a consistent analysis of the elastic neutron and proton scattering from $^{208}$Pb
and the charge exchange \pnP reaction to study the isovector part of the \nA OP,
and to estimate the Coulomb correction term (\ref{e7}) to the proton OP.

\section{Experimental method}
\label{sec2}

The measurements were performed using the MSU beam swinger time-of-flight system
\cite{Bho77,DeV79,DeV83} as modified for neutron scattering. Neutrons produced by the
$^7$Li$(p,n)^7$Be(g.s. + 0.429 MeV) reaction scatter from a cylindrical 200.64 gram target
(2.40 cm diameter x 3.90 cm long) of isotopically enriched $^{208}$Pb (98.69\%) \cite{PbT}
and are detected in a liquid scintillation counter with an overall time resolution of about
1.0 nsec. This corresponds to an energy resolution for the elastic peak of better than 1.1
MeV FWHM, sufficient to resolve the first excited state of $^{208}$Pb at 2.61 MeV. Pulse
shape discrimination is utilized to eliminate the $\gamma$-ray background. The neutron
detectors are situated in a room separated from the swinger vault by a 1.8 m thick concrete
wall, except for a hole to transmit the target scattered neutrons. Additional shielding
against neutrons coming directly from the production reaction is provided by a movable 1.1 m
long iron shadow bar. A monitor time-of-flight detector is mounted rigidly to the beam
swinger so as to measure neutron flux from the production reaction at a fixed angle near
$22^\circ$. This monitor is used to normalize the flux from run to run. Air scattering
background is accounted for by measuring target-in and target-out spectra at each angle; a
small correction is made to account for the fact that some of those air-scattered neutrons
originating behind the sample are absorbed by the sample on their way to the detector
\cite{DeV83}. Relative uncertainties are typically 2-4\% but reach 8\% at a few angles.
Observation of the $^7$Li$(p,n)$ flux at $0^\circ$ measures the product of incident neutron
flux and detector efficiency, and yields the absolute normalization to within ±3\%.
Corrections are made for the dead time, source anisotropy and background attenuation due to
the sample. Further details of the experimental procedures can be found in
Refs.~\cite{DeV79,DeV83}.

\section{Data analysis and discussion}
\label{sec3}
 A  relatively large target was used to obtain adequate statistical accuracy, making it 
necessary to correct the experimental data for the effects of multiple scattering, angle 
averaging and attenuation. Because the cross sections varied rapidly with angle we were 
concerned that the deconvolution procedures generally employed would lead to ambiguities and 
unacceptably large uncertainties. We elected to avoid these uncertainties by convoluting the 
results of optical model predictions, a straightforward procedure, before comparing them with 
the data in a search routine.  For this purpose, Kinney's finite geometry code \cite{kin70} 
was incorporated as  a subroutine of the optical model search code GIBELUMP \cite{GIB}. 
The spin-orbit part of the neutron OP was fixed at that used in Ref.~\cite{Rap78} for lower 
energy neutron data, and the``average" geometry of the WS potential was taken from
Ref.~\cite{vanO74}.

\begin{figure}[bht] \hspace*{0.5cm}
\includegraphics[width=\textwidth]{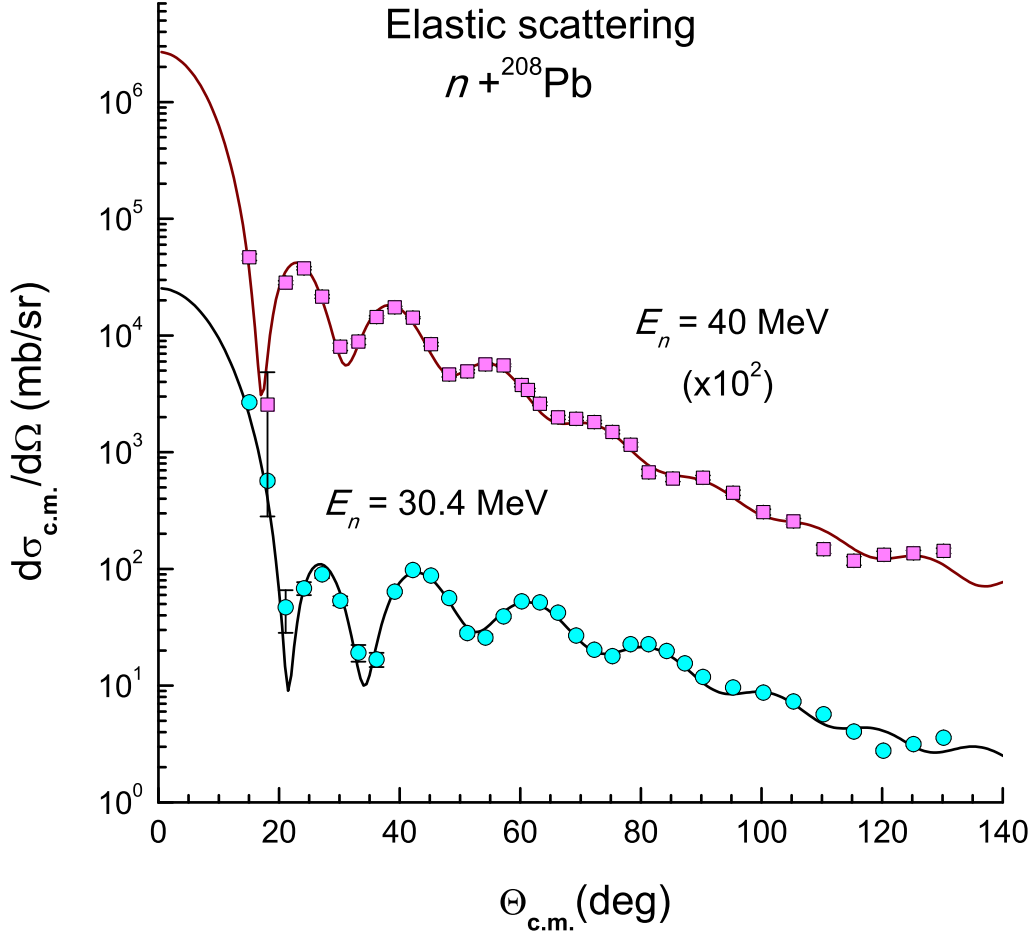}\vspace*{-0.5cm}
\caption{(Color online) Elastic \nPb scattering data at 30.4 and 40 MeV, corrected for finite geometry effects as described in the text, are
compared with the OM fit given by the modified CH89 optical potential (see OP
parameters in Table~\ref{t2}), and represent our best fit to the neutron data.}
\label{f1}\end{figure}

During the search, the smeared optical model cross sections were compared with the experimental 
data and the optical model parameters were adjusted until a good fit to the experimental data 
was obtained, thereby fixing the optical model parameters. The cross sections and their 
uncertainties resulting from this procedure were then deduced. These cross sections are 
corrected for multiple scattering, angle averaging, and attenuation, and are shown in 
Fig.~\ref{f1}, Fig.~\ref{f3} and Table~\ref{t1}. The uncertainties resulting from the finite 
geometry corrections varied with angle, and have  a maximum of about 8\% at the first 
diffraction minimum.  These uncertainties are included in the tabulated cross sections.  Because 
this is an unconventional procedure its results were compared with the normal deconvolution 
procedure used at lower energies for the case of Fe at 26 MeV; the two procedures yielded close 
agreement. The analysis process is described in much more detail in Refs.~\cite{DeV79,DeV83}.

The deduced c.m. cross sections \cite{DeV79} of the elastic \nPb scattering at 30.4 and
40 MeV have been studied in several OM analyses, including the extensive searches
for global parameters of the nucleon OP \cite{Kon03}. In the present work,
a detailed OM analysis of the measured elastic \nPb scattering data was made using
the CH89 geometric parameters of the nucleon OP \cite{Va91}. The OM analysis and
coupled-channel calculation of \pn reaction were made using the code ECIS97 written
by Raynal \cite{Ra97}

\begin{table*}
\caption{Measured cross sections of $^{208}$Pb$(n,n)^{208}$Pb at 30.4
 and 40 MeV. The cross sections have been corrected for finite geometry effects as described in the text.}\label{t1}
\begin{tabular}{cccccc|cccccccc} \hline
 & &  30.4 MeV & &  &  &  & & & & 40.0 MeV & & & \\ \hline
$\theta_{\rm c.m}$ & $\sigma_{\rm c.m}$ & $\Delta\sigma_{\rm c.m}$ &	
$\theta_{\rm c.m}$ & $\sigma_{\rm c.m}$	& $\Delta\sigma_{\rm c.m}$ & &	
$\theta_{\rm c.m}$ & $\sigma_{\rm c.m}$	& $\Delta\sigma_{\rm c.m}$ &	
$\theta_{\rm c.m}$ & $\sigma_{\rm c.m}$	& $\Delta\sigma_{\rm c.m}$ \\
(deg)	&	(mb/sr)	&	(mb/sr)	&	(deg)	&	(mb/sr)	&	(mb/sr)	& &	(deg)	&	
(mb/sr)	&	(mb/sr)	&	(deg)	&	(mb/sr)	&	(mb/sr)	\\ \hline
15.07	&	2675	&	83.0	&	66.26	&	42.24	&	1.13	& &	15.07	&	468.6	&	
30.5	&	63.26	&	26.01	&	0.92	\\
18.09	&	568.5	&	33.2	&	69.27	&	26.92	&	0.83	& &	18.09	&	25.62	&	
22.8	&	66.26	&	19.89	&	0.79	\\
21.1	&	47.02	&	18.7	&	72.27	&	20.33	&	0.67	& &	21.1	&	283.7	&	
10.9	&	69.27	&	19.32	&	0.65	\\
24.12	&	68.27	&	8.7	&	75.28	&	17.95	&	0.61	& &	24.12	&	376.7	&	
11.2	&	72.28	&	18.11	&	0.58	\\
27.13	&	89.94	&	6.4	&	78.28	&	22.67	&	0.61	& &	27.13	&	215.8	&	
7.6	&	75.28	&	14.93	&	0.56	\\
30.14	&	53.39	&	4.75	&	81.28	&	22.56	&	0.58	& &	30.14	&	80.11	&	
5.9	&	78.28	&	11.6	&	0.52	\\
33.16	&	19.2	&	3.15	&	84.29	&	19.83	&	0.56	& &	33.16	&	88.84	&	
5.1	&	81.29	&	6.722	&	0.436	\\
36.17	&	16.79	&	2.33	&	87.29	&	15.57	&	0.455	& &	36.17	&	143.8	&	
4.04	&	85.29	&	5.942	&	0.341	\\
39.18	&	63.85	&	2.19	&	90.29	&	11.89	&	0.435	& &	39.18	&	174.1	&	
3.9	&	90.29	&	6.058	&	0.274	\\
42.19	&	97.82	&	2.22	&	95.29	&	9.654	&	0.299	& &	42.19	&	142.2	&	
2.49	&	95.29	&	4.5	&	0.237	\\
45.2	&	87.86	&	1.52	&	100.3	&	8.721	&	0.261	& &	45.2	&	84.29	&	
2.91	&	100.3	&	3.072	&	0.178	\\
48.21	&	56.29	&	1.37	&	105.3	&	7.321	&	0.236	& &	48.22	&	46.44	&	
2.28	&	105.3	&	2.558	&	0.145	\\
51.22	&	28.2	&	1.84	&	110.3	&	5.704	&	0.194	& &	51.23	&	49.27	&	
1.6	&	110.3	&	1.477	&	0.123	\\
54.23	&	25.85	&	1.68	&	115.3	&	4.024	&	0.164	& &	54.23	&	56.53	&	
1.53	&	115.3	&	1.176	&	0.10	\\
57.24	&	39.21	&	1.66	&	120.2	&	2.772	&	0.137	& &	57.24	&	55.5	&	
1.46	&	120.3	&	1.325	&	0.09	\\
60.25	&	52.85	&	1.15	&	125.2	&	3.164	&	0.136	& &	60.25	&	37.61	&	
1.27	&	125.2	&	1.362	&	0.081	\\
63.26	&	51.68	&	1.11	&	130.2	&	3.584	&	0.093	& &	61.25	&	34.2	&	
1.01	&	130.2	&	1.432	&	0.077	\\ \hline
\end{tabular}
\end{table*}

\subsection{Coulomb correction to the proton incident energy}
\begin{figure}[bht] \vspace*{-1cm}
\includegraphics[width=\textwidth]{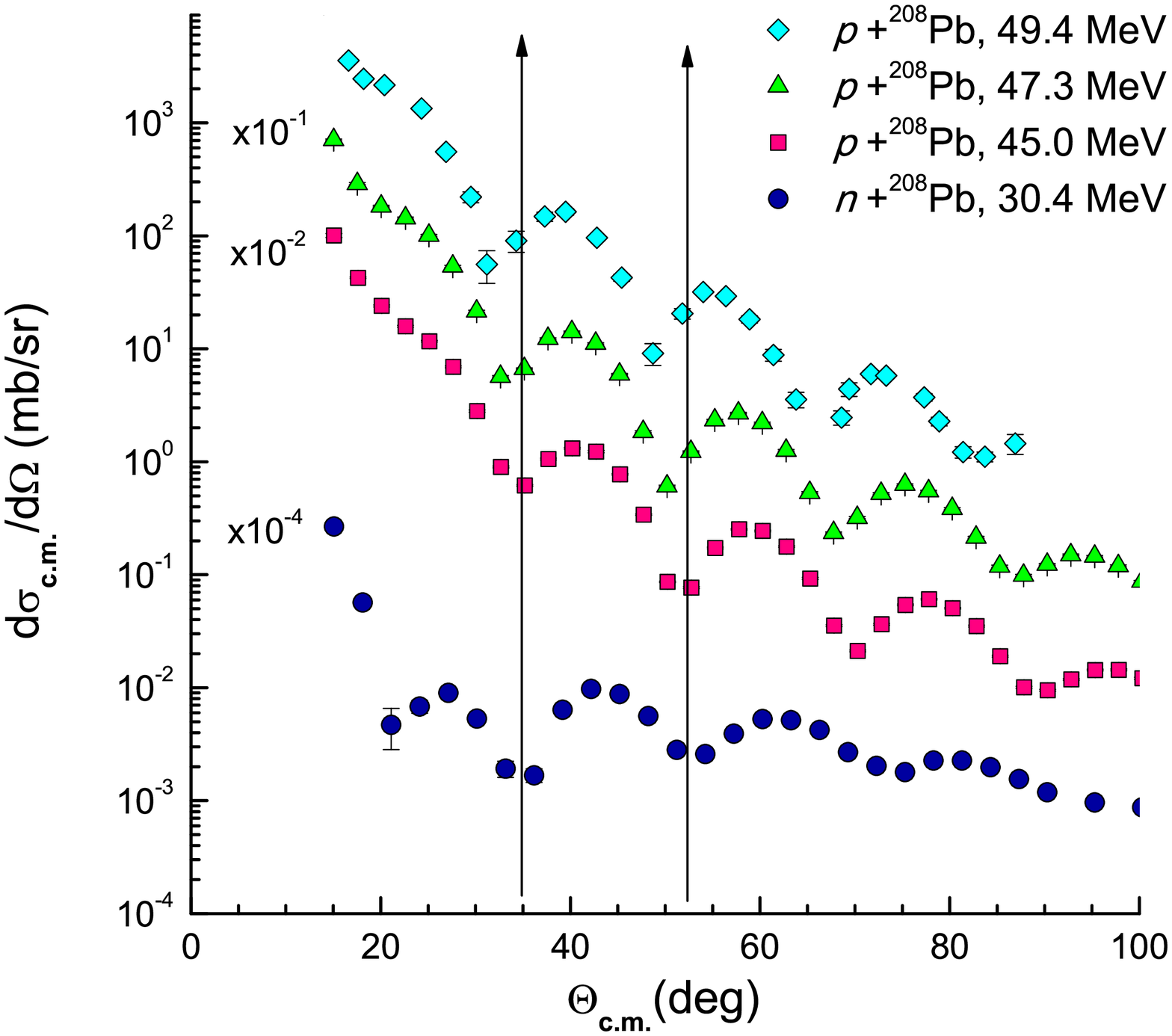}\vspace*{-1cm}
 \caption{(Color online) Diffractive oscillation of the elastic \nPb scattering
 data at 30.4 MeV in comparison with those of elastic \pPb scattering data at
 45, 47.3 \cite{vanO74} and 49.4 MeV \cite{Man71}.} \label{f2}
\end{figure}

The Coulomb correction to the proton incident energy arises because
the proton slows down in the repulsive Coulomb field of the
nucleus and, hence (because the real part of the OP decreases with increasing
energy), the real part of $U_p$ is more attractive compared to that of the
neutron OP at the same bombarding energy, even when $U_1 = 0$. Estimation of the
Coulomb correction of the nucleon OP requires that one first determine the difference
in the effective proton and neutron incident energies, so that the same
isoscalar and isovector potentials $U_{0(1)}$ can be used to generate the proton
and neutron OPs using Eq.~(\ref{e3a}).


The difference in the proton and neutron energies has usually been assumed
either to be constant at the average Coulomb energy of the incident proton,
$\Delta E_C=6Ze^2/(5R_C)\approx 19$ MeV for a $^{208}$Pb target \cite{Va91},
or to be energy dependent \cite{Kon03}. In the present work we have chosen
the CH89 global parametrization for the nucleon OP, which has a simple
functional form and is quite reliable for the nucleon elastic scattering
from medium-mass nuclei at energies of 10 to 65 MeV, as the starting point
of our OM analysis. The CH89 complex nucleon OP is determined as the following
energy dependent functional
\begin{equation}
 U(E)=U_0(E-\Delta E_C)\mp\varepsilon U_1(E-\Delta E_C), \label{e7b}
\end{equation}
with - and + sign pertaining to the incident proton and neutron, respectively.
Thus, the CH89 proton and neutron optical potentials determined at energies
$E_p$ and $E_n=E_p-\Delta E_C$, respectively, should be fully consistent
with the Lane formalism (\ref{e3a})-(\ref{e3b}). We discuss below the extent
to which this assumption accurately describes the experimental data.

To obtain an estimate of  $\Delta E_C$ based on experiment, we have determined
the proton bombarding energy at which the slowed proton and the neutron
have the same average momentum, and thereby the diffraction maxima and minima
of the proton and neutron angular distributions fall at about the same (average)
angles in the forward region \cite{DeV81}. For this purpose, the diffractive
oscillation of the elastic \nPb data at 30.4 MeV has been compared with those
of the elastic \pPb scattering at 45, 47.3  and 49.4 MeV (see Fig.~\ref{f2}).
It is clear that the oscillation pattern of the 30.4 MeV neutron data does not
agree with that of the 49 MeV proton data, as would be expected from the average
Coulomb energy prescription of the CH89 model. The elastic \pPb scattering data
at 45 MeV \cite{vanO74} do have an oscillation pattern at forward
angles very similar to that of the 30.4 MeV neutron data.  This implies that
$\Delta E_C$ $\thickapprox 14.6$ MeV, significantly smaller than the value
of 19 MeV used in the CH89 parametrization.

Our procedure presumably equalizes the average momenta
or wavelengths of the neutrons and protons in the regions dominating the scattering.
\begin{figure}[bht] \vspace*{0cm}
\includegraphics[width=\textwidth]{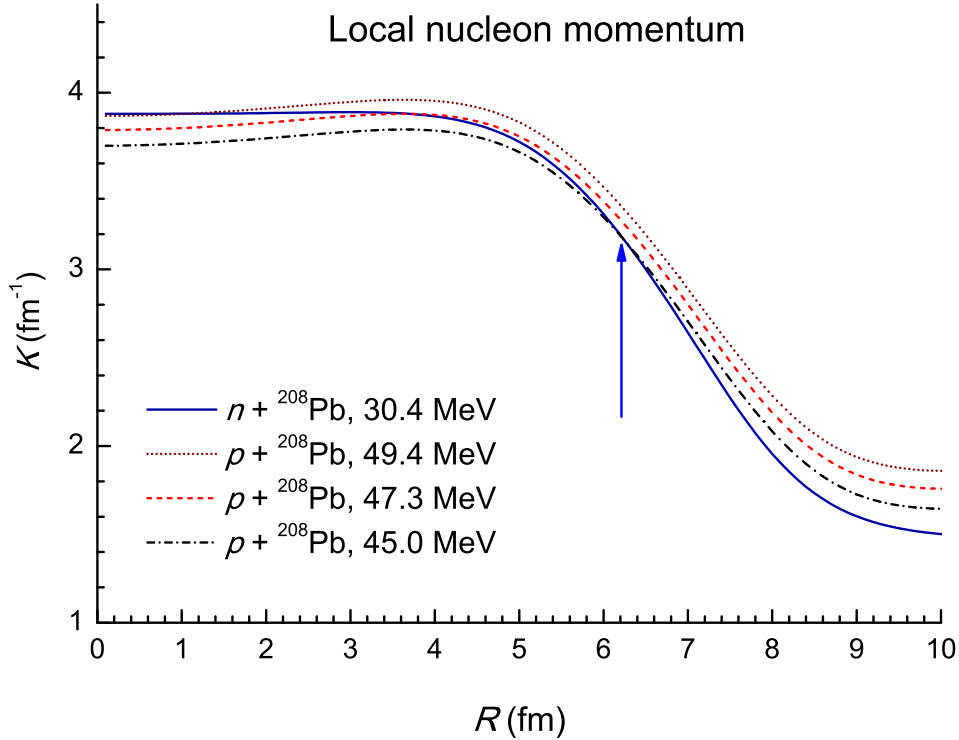}
 \caption{(Color online) Local nucleon momentum $K(R)$ determined from
 Eq.~(\ref{e8}) with the real OP given by the folding calculation
 \cite{Kho07,Kho02}, using the density dependent CDM3Y6 interaction \cite{Kho97}.}
\label{f4} \end{figure}
For illustration, we have plotted in Fig.~\ref{f4} the local nucleon (relative motion)
momentum $K(R)$ determined from the real folded nucleon OP \cite{Kho02} as
\begin{equation}
 K^2(E,R)=\frac{2\mu}{{\hbar}^2}[E_{\rm c.m.}-V(R)-V_{\rm C}(R)],
\label{e8}
\end{equation}
where $\mu$ is the nucleon reduced mass, $V(R)$ and $V_{\rm C}(R)$ are,
respectively, the real central nuclear and Coulomb parts of the OP
($V_{\rm C}\equiv 0$ for the neutron-nucleus system). It can be seen from
Fig.~\ref{f4} that the local momentum of 30 MeV neutrons is equal that of
45 MeV protons at $R\sim 6$ fm, in the surface region of the $^{208}$Pb
target. Because the diffractive scattering is dominantly determined by the
strength and shape of the OP at the surface, the results plotted in
Fig.~\ref{f4} support a Coulomb correction of $\Delta E_C\approx
14.6$ MeV at a proton incident energy of 45 MeV.

\begin{figure}[bht] \vspace*{-1cm}
\includegraphics[width=\textwidth]{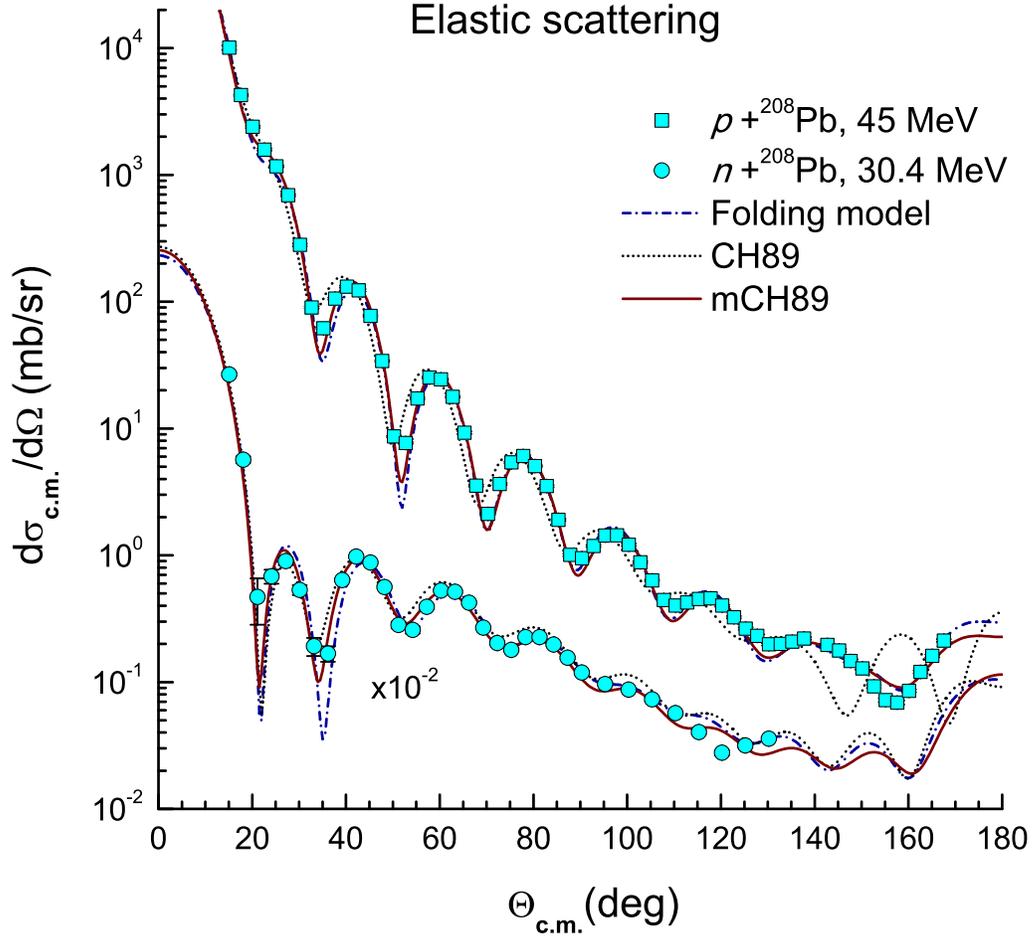}\vspace*{-1cm}
 \caption{(Color online) OM description of the elastic \nPb data at 30.4 MeV
and \pPb data at 45 MeV \cite{vanO74} given by the CH89 (dotted line),
mCH89 (solid line), and the folded (dash-dotted line) optical
potentials.}  \label{f3}
\end{figure}

Finally, in Fig.~\ref{f3} we show OM calculations with several related optical
potentials to assess  differences in the cross sections they predict. First, the
two data sets were compared with calculations using the original CH89 parameters
\cite{Va91}, but with  $\Delta E_C$ set to 14.6 MeV at 45 MeV proton energy.
Then the potential depths were adjusted, yielding mCH89 OP, to give the best
$\chi^2$ fit to the data. For a comparison, OM results given by the complex
folded potential \cite{Kho07} calculated with the CDM3Y6 interaction \cite{Kho97}
(strengths of the real and imaginary folded potentials were adjusted to the
 best $\chi^2$ fit to the data) are shown.

\subsection{\pnP data and the isovector term of the OP}
\begin{figure}[bht] \vspace*{-0.5cm}
\includegraphics[width=\textwidth]{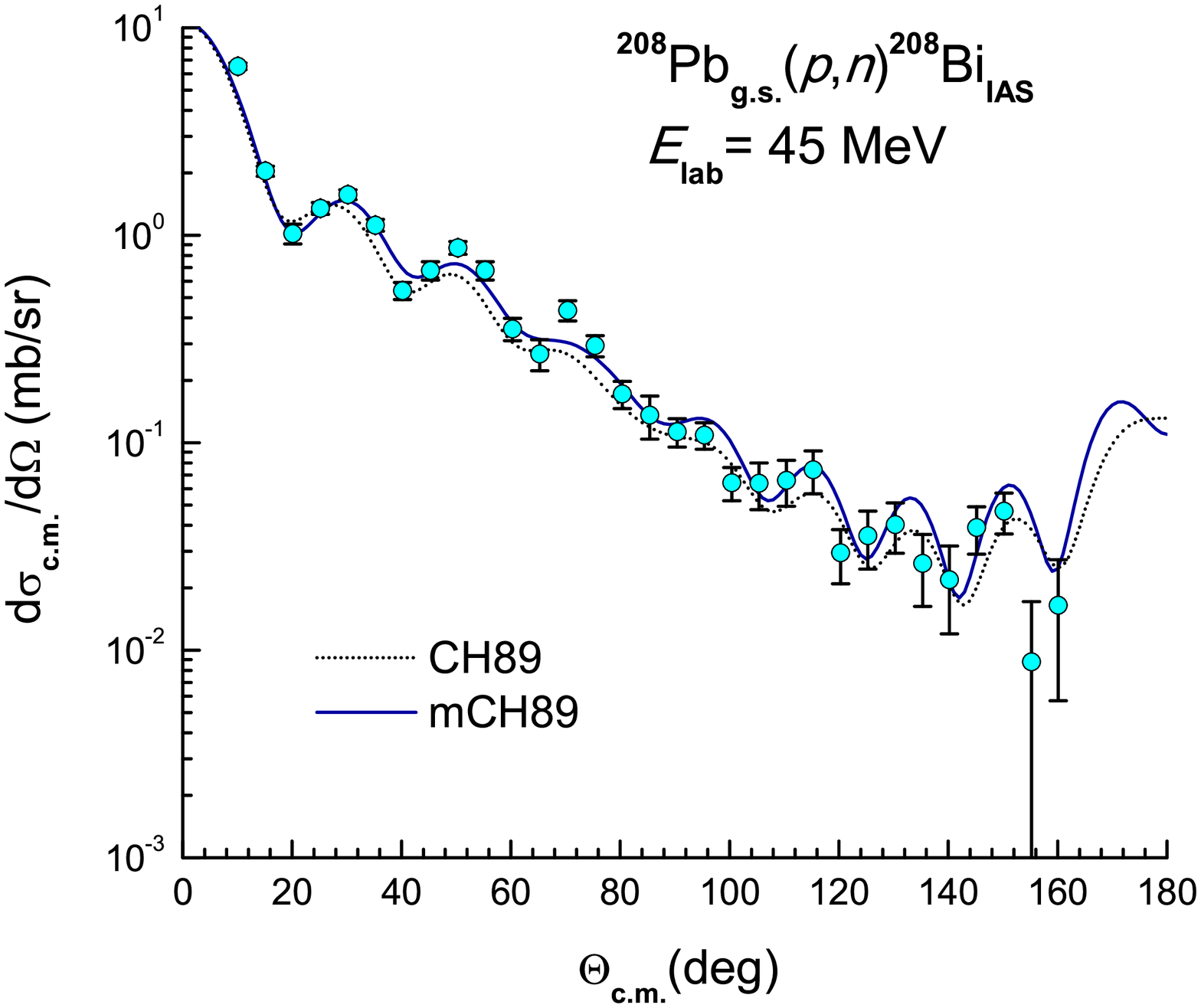}\vspace*{-0.5cm}
 \caption{(Color online) CC description of the charge exchange \pnP data
 measured at 45 MeV \cite{Doe75} given by $F_{pn}$ built upon
 the isovector part $U_1$ of the CH89 nucleon OP \cite{Va91}. The two
 different curves were obtained with two choices of the 45 MeV proton OP
 (see Table~\ref{t2})} \label{f5}
\end{figure}
In the two-channel approximation for the charge exchange \pn reaction
exciting IAS, the total wave function can be written as
\begin{equation}
\Psi=|pA\rangle\chi_{pA}({\bm R})+|n\tilde{A}\rangle\chi_{n\tilde{A}}({\bm R}),
 \label{e8a}
\end{equation}
where the $\chi({\bm R})$ describe the relative \nA motion. The elastic
$(p,p)$ scattering and charge exchange \pnA cross sections are then obtained
from the solutions of the following coupled-channel (CC) equations \cite{Sat83}
\begin{eqnarray}
\left[K_p+U_p(R)-E_p\right]
 \chi_{pA}({\bm R})=-F_{pn}(R)\chi_{n\tilde{A}}({\bm R}),
  \label{e9a} \\
\left[K_n+U_n(R)-E_n\right]
 \chi_{n\tilde{A}}({\bm R})=-F_{pn}(R)\chi_{pA}({\bm R}).
 \label{e9b}
\end{eqnarray}
Here $K_{p(n)}$ and $E_{p(n)}$ are the kinetic-energy operators and
c.m. energies of the \Ap and \An partitions. The proton OP in the entrance
(\Ap) channel was determined from the best OM fit to the elastic \pPb scattering
data at 45 MeV (see Fig.~\ref{f3} and potential parameters in Table II), while
the neutron OP in the outgoing (\An) channel was constructed from the
isoscalar and isovector parts of the nucleon OP using the standard isospin
coupling scheme \cite{Sat83,Kho07}. Since the energies of isobar analog
states are separated approximately by the Coulomb displacement energy,
the \pn transition between them has a nonzero $Q$ value. To account for
this effect, the isoscalar $U_0$ and isovector $U_1$ potentials used
to construct $F_{pn}(R)$ and $U_n(R)$ were evaluated from the CH89
systematics at an effective incident energy of $E=E_{\rm lab}-Q/2$, midway
between the energies of the incident proton and emergent neutron \cite{Sat64}.
Given the elastic \pPb scattering
data \cite{vanO74} and charge exchange \pnP data \cite{Doe75} (both measured
at 45 MeV), we were able to test the isovector term $U_1$ of the proton OP
by comparing the results of the CC calculation for the
\pn cross section with the data. One can see in Fig.~\ref{f5}
that the isovector part of the CH89 nucleon OP \cite{Va91} accounts very
well for the \pn data, especially when the WS strengths of the proton OP
are optimized by the best $\chi^2$ fit to the elastic \pp data at 45 MeV
(the mCH89 potential in Table~\ref{t2}). This result shows that the complex
isovector potential $U_1$ given by the CH89 parametrization for the proton OP
at 45 MeV can be used to estimate the Coulomb correction using Eq.~(\ref{e7})
and the corresponding neutron optical potential $U_n$.

\subsection{Coulomb correction to the proton OP}
After $\Delta E_C$ has been fixed to give
the same diffraction patterns at the forward angles for the proton and
neutron elastic cross sections at $E_p$ and $E_n=E_p-\Delta E_C$,
respectively, one might naively expect from Eq.~(\ref{e7b}) that the
corresponding proton and neutron optical potentials are fully Lane consistent.
But this is only true if the only physics involved is that due to the energy
shift. Yet we know there are other effects, mostly affecting the imaginary
potential, for example, Coulomb excitation, different Q values for
\pn and $(n,p)$ reactions, and different level structures. In the following
we obtain an estimate of the importance of such phenomena.

Given that  the isovector potential $U_1$ of the CH89 potential was shown
above to give a realistic description of the \pn data, it is reasonable to use
$U_1$ given by the CH89 systematics to estimate $\Delta U_C$, based on the
OM analysis of the elastic \pPb and \nPb scattering data measured at 45
and 30.4 MeV. In our OM analysis we used the same WS functional form for the
nucleon OP as that used by the CH89 systematics \cite{Va91}, so the real and
imaginary parts of the nucleon OP are determined as
\begin{eqnarray}
 V(R)&=&-V_vf_{\rm ws}(R,R_v,a_v)+\frac{2V_{\rm so}}{R}
 \frac{d}{dR}f_{\rm ws}(R,R_{\rm so},a_{\rm so})({\bm l}.{\bm \sigma}) ,
  \label{e10a} \\
 W(R)&=&-W_vf_{\rm ws}(R,R_w,a_w)+4a_wW_s
 \frac{d}{dR}f_{\rm ws}(R,R_w,a_w),  \label{e10b} \\
 {\rm where} & & f_{\rm ws}(R,R_x,a_x)=1/\{1+\exp[(R-R_x)/a_x]\}. \nonumber
 \end{eqnarray}
The real part of the proton OP includes the Coulomb potential $V_C(R)$
taken from the CH89 systematics \cite{Va91}. The CH89 parametrization gives
the isovector part $U_1$ of the nucleon OP as
\begin{equation}
 U_1(R)=V_1f_{\rm ws}(R,R_v,a_v)-i4a_wW_1 \frac{d}{dR}f_{\rm ws}(R,R_w,a_w).
 \label{e10c}
\end{equation}
For the proton OP at 45 MeV, the complex strength of the isovector potential is
readily obtained from the CH89 parametrization as $\varepsilon V_1=2.77 \pm 0.17$
MeV and $\varepsilon W_1=2.05 \pm 0.13$ MeV. The quoted uncertainties
were determined from the systematic uncertainties of
the CH89 global parameters \cite{Va91}.
\begin{table*}
\caption{OP parameters (\ref{e10a})-(\ref{e10b}) used in the OM analysis of the
elastic \pPb data and \nPb data. The radii and diffuseness parameters were
given by the CH89 parametrization (in fm): $R_v=7.18,\ R_w=7.46,\ a_v=a_w=0.69,\
R_{\rm so}=6.73,\ a_{\rm so}=0.63$. The incident energy $E$, Coulomb corrections
(\ref{e11a})-(\ref{e11b}) and potential depths are given in MeV. The real strength of
the isovector potential was taken as $\varepsilon V_1 = 2.77 \pm 0.17$ MeV, its imaginary
strength $\varepsilon W_1$ and the Coulomb correction strengths $\Delta V_v,\ \Delta W_v$,
and $\Delta W_s$ are given with uncertainties (in brackets) estimated from those
of our OM fit and the CH89 potential parameters \cite{Va91}.}\vspace*{1cm}
\label{t2}
\begin{tabular}{|c|c|c|c|c|c|c|c|c|c|c|}
\hline
System & $E$ & Potential & $V_v$ & $\Delta V_v$ & $W_v$ & $\Delta W_v$ &
 $W_s$ & $\Delta W_s$  & $\varepsilon W_1$ & $V_{\rm so}$ \\
\hline
\pPb & 54.2 & CH89 & 43.71 & 0.00 & 4.50 & 0.00 & 6.53 & 0.00 & -1.80 (0.11) & 5.90 \\
   & & mCH89 & 40.61 & -1.90 (0.56) & 4.67 & -0.20 (0.30) & 5.03 & -1.88 (0.51)
   & -1.80 (0.11) & 5.90 \\ \hline
\nPb  & 40.0 & CH89  & 38.17 & - & 4.50 & -  & 2.93  & -  & 1.80 (0.11) & 5.90  \\
  &  & mCH89   & 36.97 & -   & 4.87 & -  & 3.31 & -  & 1.80 (0.11) & 6.11 \\\hline
\pPb  & 45.0 & CH89 & 46.58  &  0.00 & 3.34 & 0.00 & 7.43 & 0.00 & -2.05 (0.13) & 5.90 \\
  &  &mCH89  & 43.27 & -1.94 (0.54) & 3.69 & -0.75 (0.26)  & 5.72  & -1.18 (0.52)
  & -2.05 (0.13) & 4.22  \\ \hline
\nPb  & 30.4 & CH89 & 41.04  & - & 3.34 & - & 3.33 & - & 2.05 (0.13) & 5.90  \\
 &  & mCH89 & 39.67 & - & 4.44 & - & 2.80 & - & 2.05 (0.13) & 6.47 \\ \hline
\end{tabular}
\end{table*}

\begin{figure}[bht]\vspace*{-1cm}
\includegraphics[width=\textwidth]{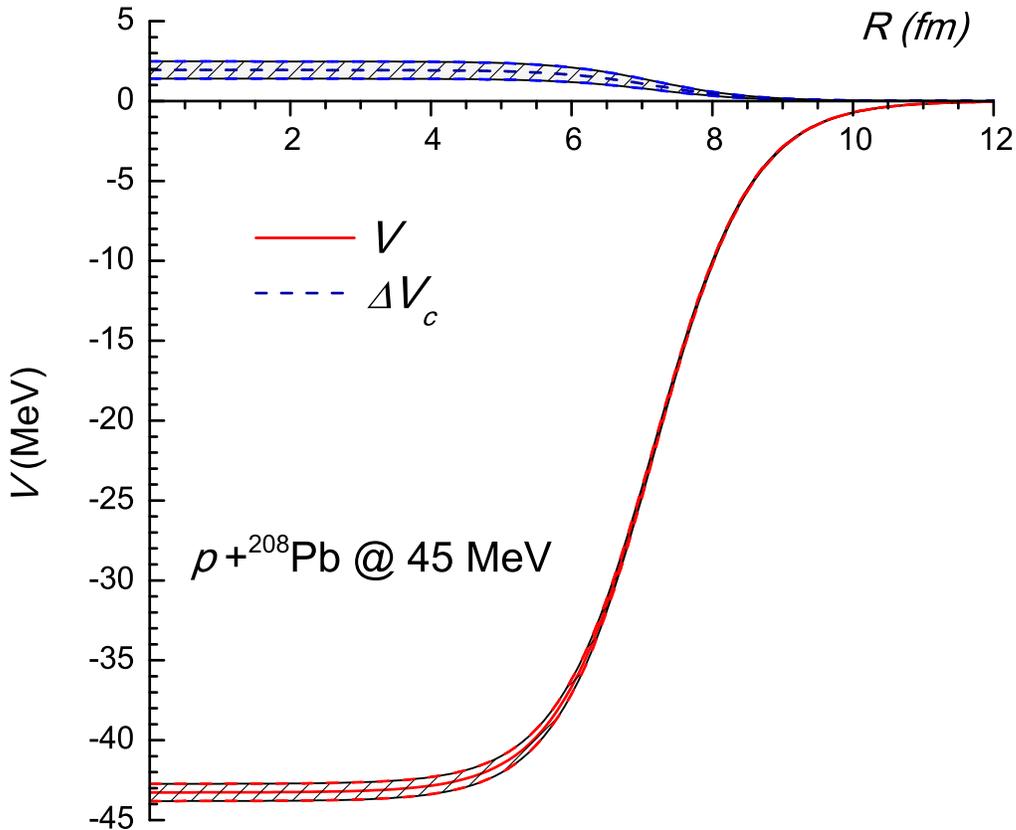}\vspace*{-2cm}
 \caption{(Color online) Real central OP of 45 MeV protons and the
corresponding Coulomb correction $\Delta V_C$ determined from
Eq.~(\ref{e11a}) using the mCH89 optical potentials for 30 MeV neutrons
and 45 MeV protons. The uncertainties were estimated from that given by the CH89
systematics for $\varepsilon\Delta V_1$ and standard errors of the OM
fit with the code ECIS97 \cite{Ra97}.} \label{f6}
\end{figure}
\begin{figure}[bht]\vspace*{-1cm}
\includegraphics[width=\textwidth]{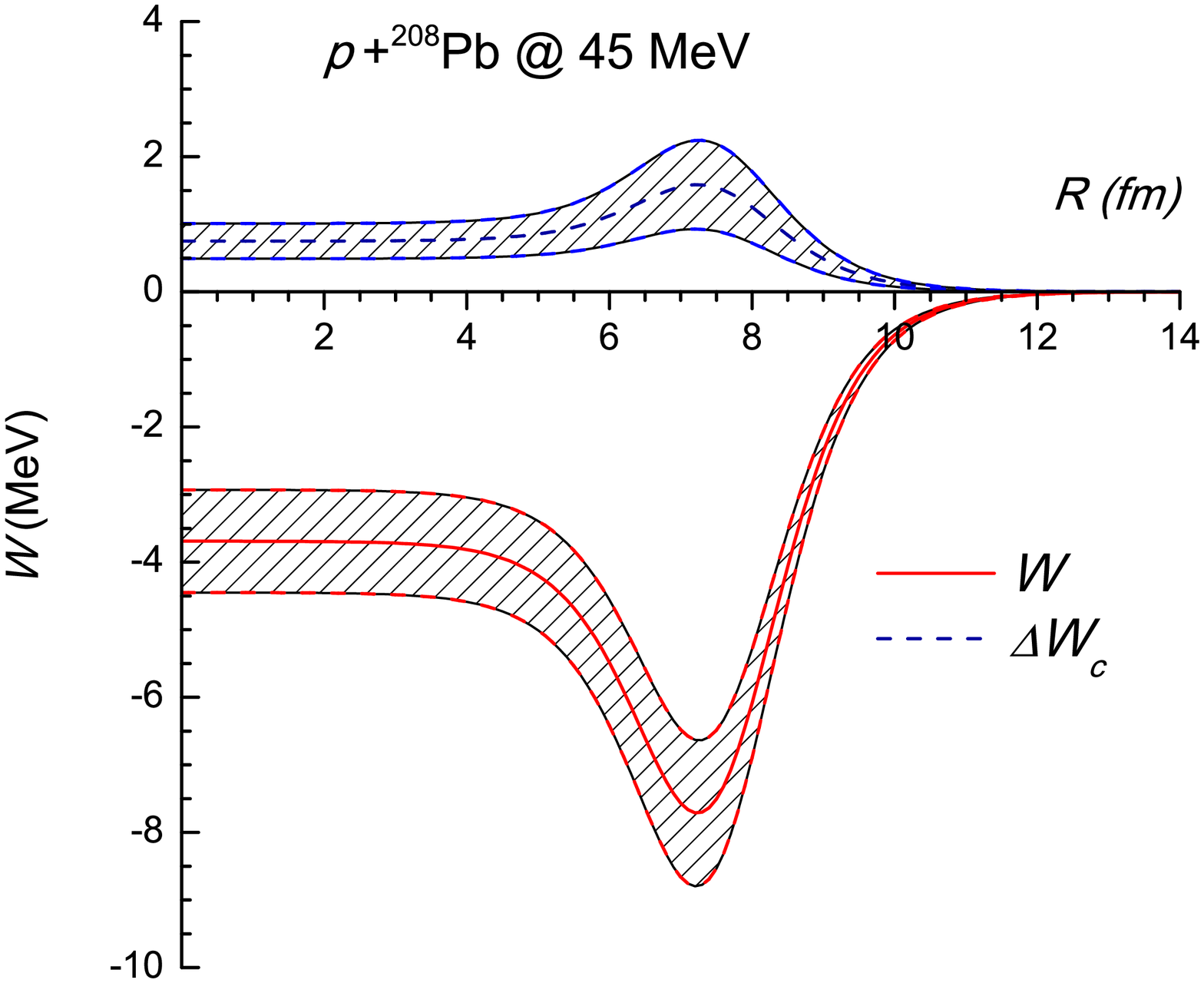}\vspace*{-2cm}
\caption{(Color online) The same as Fig.~\ref{f6} but for the imaginary part
of the central OP, with the corresponding Coulomb correction $\Delta W_C$
determined from Eq.~(\ref{e11b}).} \label{f7}
\end{figure}
In the present OM analysis, we first generated the complex \pPb and \nPb optical
potentials using the CH89 parametrization but with a corrected $\Delta E_C=14.6$ MeV
for the proton OP (see CH89 parameters in Table~\ref{t2}). As a result, the CH89
optical potentials for 45 MeV proton and 30.4 MeV neutron are
\emph{fully Lane consistent}, with the Coulomb correction taken into
account only by using the energy shift $\Delta E_C$ in Eq.~(\ref{e7b}), and its
remnant, determined by Eqs.~(\ref{e11a})-(\ref{e11b}), is exactly zero.
The OM description of the 30 MeV neutron- and 45 MeV proton elastic scattering
by the Lane consistent CH89 optical potentials is shown in Fig.~\ref{f3}
as dotted line. One can see that the Lane consistent CH89 OP describes
the data fairly well, excepting at the backward angles where it fails to follow
the oscillation pattern of the measured $(p,p)$ data. A much improved OM description of
the data has been achieved by adjusting the depths of the OP while keeping the same WS
geometry as that of the CH89 potential (see mCH89 parameters in Table~\ref{t2}
and solid lines in Fig.~\ref{f3}). Such an adjustment procedure naturally gave rise
to a non-zero remnant of the Coulomb correction $\Delta U_C=\Delta V_C+i\Delta W_C$.
Assuming the same $U_1$ for the mCH89 OP as that of the original CH89 OP,
the \emph{complex} remnant of the Coulomb correction can be explicitly
determined from Eqs.~(\ref{e7}),(\ref{e10a}), and (\ref{e10b}) as
\begin{eqnarray}
 \Delta V_C(R)&=&-\Delta V_vf_{\rm ws}(R,R_v,a_v),
  \label{e11a} \\
 \Delta W_C(R)&=&-\Delta W_vf_{\rm ws}(R,R_w,a_w)+4a_w\Delta W_s
 \frac{d}{dR}f_{\rm ws}(R,R_w,a_w).  \label{e11b}
 \end{eqnarray}
The deduced WS strengths (\ref{e11b}) of $\Delta U_C$, further referred to as the
Coulomb correction to the mCH89 proton OP, are given in Table~\ref{t2} with the
uncertainties estimated consistently from the standard errors of the OM fits and
those of the CH89 potential parameters.
Although the strengths of the spin-orbit potential $V_{\rm so}$ of the best
fit mCH89 optical potentials also differ from the original CH89 values,
we did not assign this difference to the isospin impurity of the OP caused by
the Coulomb correction.

From the strengths of the Coulomb correction $\Delta U_C$ to the best-fit mCH89
optical potentials for the elastic neutron and proton scattering at 30.4 MeV
and 45 MeV, respectively, we find following ratios of the Coulomb correction to
the strength of the OP: $\Delta V_v/V_v\approx 4.5\%,\ \Delta W_v/W_v\approx 20.3\%,$
and $\Delta W_s/W_s\approx 20.6\%$. It is obvious that, within the Lane formalism
(\ref{e1}), these ratios give us a realistic estimate of the \emph{isospin impurity}
of the mCH89 nucleon OP. The radial dependence of the Coulomb correction to the real
and imaginary parts of the central OP is plotted in Figs.~\ref{f6} and ~\ref{f7},
respectively, where one can see clearly a repulsive character of the Coulomb
correction to both the real and imaginary OP. In terms of the isospin impurity,
we conclude that the Lane formulation (\ref{e1}) is accurate to within
about 4-5\% for the real central OP. However, the isospin impurity becomes
much larger (above 20\%) for the imaginary part of the CH89 OP and reaches its
peak at the surface.

\begin{figure}[bht] \vspace*{-1cm}
\includegraphics[width=\textwidth]{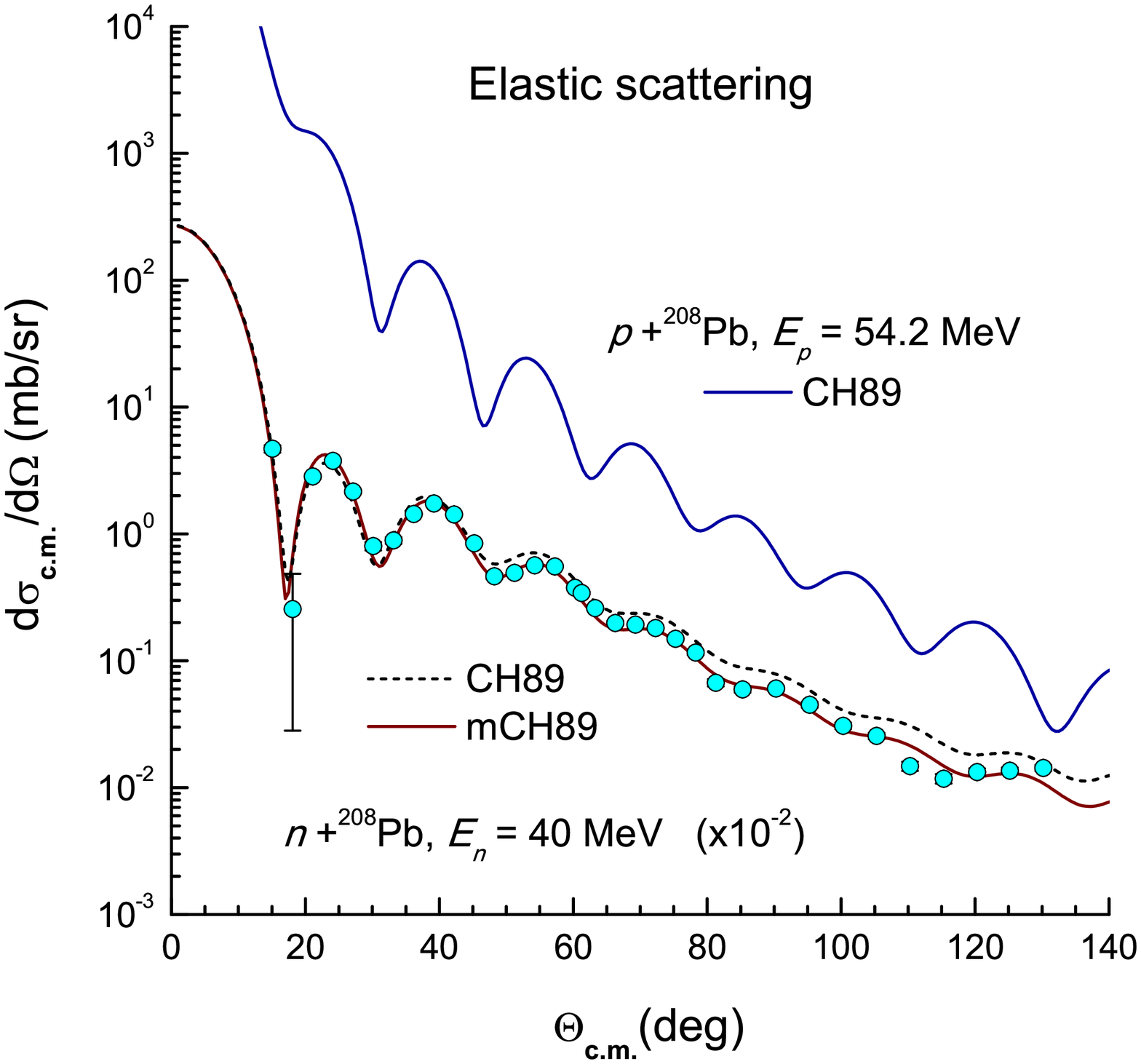}\vspace*{-0.5cm}
 \caption{(Color online) OM description of the elastic \nPb data at 40 MeV
given by the CH89 OP (dashed line) and mCH89 OP (solid line). The OM
prediction for the elastic \pPb scattering at 54.2 MeV is given by
the CH89 OP, with the Coulomb correction to the proton incident energy
$\Delta E_C\approx 14.2$ MeV.} \label{f8}
\end{figure}
In a study of the elastic \nPb scattering at 40 MeV similar to that
described above for 30.4 MeV neutrons, and using
the complex CH89 OP \cite{Va91} to predict the elastic \pPb scattering
at the higher energies, we find that the elastic proton scattering
at 54.2 MeV has about the same forward angle diffraction pattern
as the measured elastic \nPb scattering at 40 MeV, yielding $\Delta E_C\approx 14.2$
MeV for the 54 MeV proton potential. There are no elastic proton data available
near 54 MeV, so we compare in Fig.~\ref{f8} only the OM prediction for the elastic
\pPb scattering at 54.2 MeV with the elastic \nPb data at 40 MeV. In order to
roughly estimate the Coulomb correction for 54.2 MeV protons we
have scaled the real and imaginary WS strengths of the CH89 OP by
the same factors deduced from the corresponding strengths
of the mCH89 OP compared to those of the CH89 OP for 45 MeV protons.
\begin{figure}[bht] $\vspace*{0cm}$
\includegraphics[width=\textwidth]{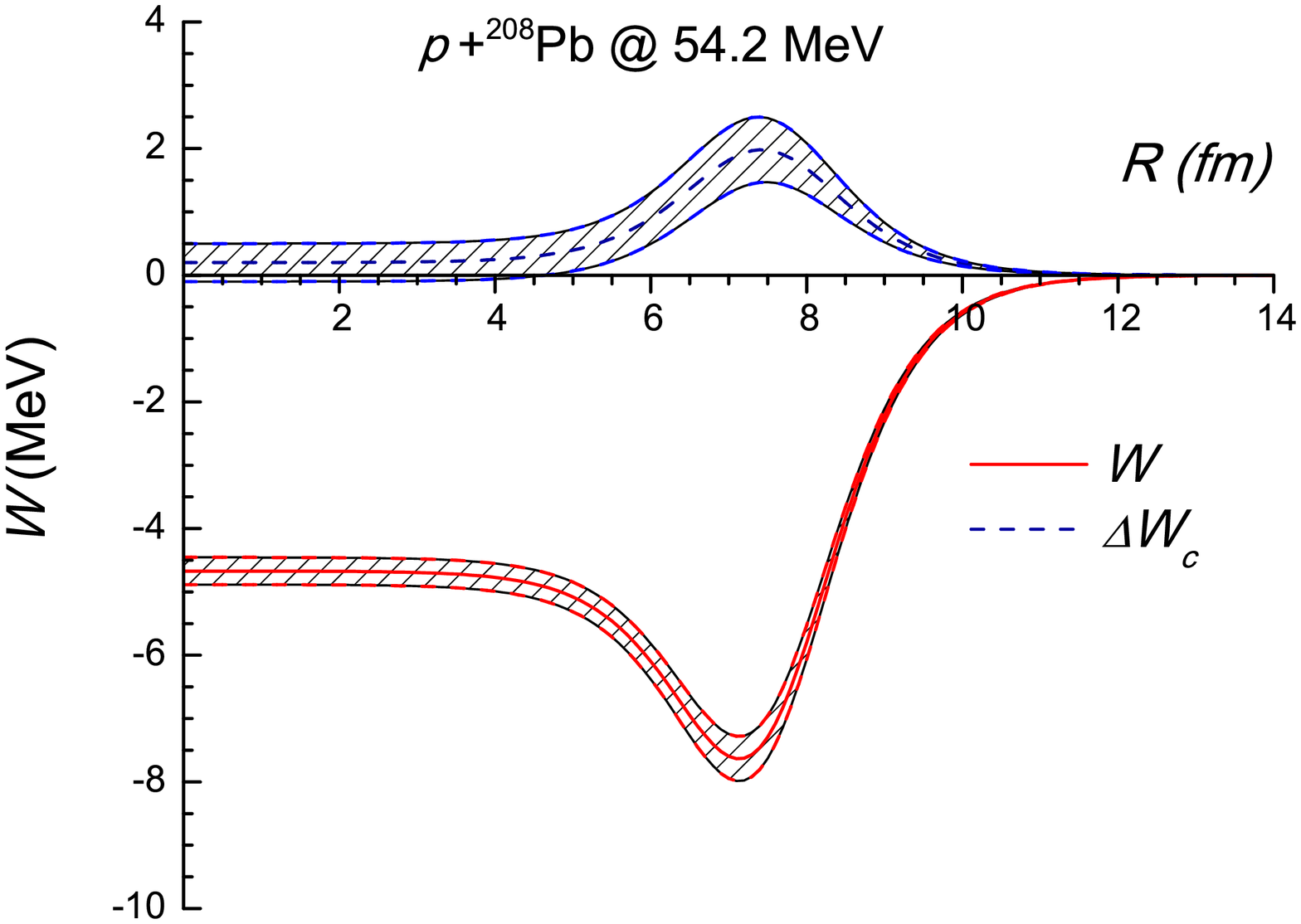}\vspace*{-2cm}
\caption{(Color online) The same as Fig.~\ref{f7} but for the imaginary
part of the mCH89 optical potential of 54.2 MeV proton (see Table~\ref{t2}).}
\label{f9}\end{figure}
The  parameters obtained are given in Table~\ref{t2} as mCH89. We note that
the imaginary parts of the CH89 and mCH89 proton OPs at 54.2 MeV give
total reaction cross sections $\sigma_R\approx 2121$ and 2050 mb, respectively,
and that the latter value agrees quite well with the experimental trend:
measured values of $\sigma_R$ lie around 2000 mb at proton energies
of 40 to 60 MeV \cite{Ca96}. We have further assumed the same $U_1$ for
the mCH89 proton OP at 54.2 MeV as that given by the original CH89
parameters for 54.2 MeV protons, and deduced the Coulomb correction
$\Delta U_C$ to the mCH89 proton OP in the same way as done above
for 45 MeV protons. We find a  behaviour of $\Delta U_C$ very similar to
that found for 45 MeV protons, with a rather strong
$\Delta W_C$ peaked at the nuclear surface (see Fig.~\ref{f9}).

\section{Discussion}

We find that \pPb data at proton incident energies of 45 and 54.2 MeV have closely
the same diffraction pattern as elastic neutron scattering at 30.4 and 40 MeV,
corresponding to energy shifts $\Delta E_C$ of 14.6 and 14.2, respectively,
owing to Coulomb repulsion effects.
Using $E_p=E_n+\Delta E_C$ (with the Coulomb correction taken into account by the
CH89 systematics)  we found that the best-fit proton OP still contains a non-zero
remnant of the Coulomb correction, which represents the isospin impurity of the
nucleon OP. While the correction to the real part of the proton OP is only a few
percent, as might be expected from the mean-field nature of the \emph{real}
nucleon OP, the correction to imaginary part at the nuclear surface is about 20\%.

Such a significant isospin impurity of the imaginary OP found in our analysis
also confirms the trend found recently in a global dispersive optical model
(DOM) analysis of elastic proton and neutron scattering \cite{Mue11}. Namely,
from a comparison of the elastic proton and neutron data on $^{208}$Pb, the
surface component of the imaginary OP obtained in the DOM has shown quite
different $(N-Z)$ asymmetry dependences between protons and neutrons. Such
a difference results directly on a deviation of the isospin dependence of the
nucleon OP from the Lane form (\ref{e3a})-(\ref{e3b}) and gives rise, therefore,
to a larger Coulomb correction to imaginary OP at the nuclear surface.

These differences indicate that there are different mechanisms for proton and
neutron absorption that are linked to different non-elastic reaction
channels induced by proton and neutron on $^{208}$Pb target. The most obvious
of these are Coulomb excitation and a difference in the $Q$ values of the
$(n,p)$ and \pn reactions. Because the physics origin of the imaginary OP is
multifaceted and contains dynamic higher-order (beyond the mean-field) contributions,
the significant isospin impurity found for the imaginary part of the nucleon OP
was not unexpected. One needs an explicit microscopic
calculation of the imaginary OP based, e.g., on the Feshbach formalism
\cite{Fe92}, to establish full Lane consistency of the mean-field
part of the nucleon OP. The present results are also in a qualitative
agreement with the Brueckner-Hartree-Fock calculation of the nucleon OP by
Jeukenne {\it et al.} \cite{Jeu77}, which showed that for heavy nuclei
such as $^{208}$Pb the imaginary Coulomb correction can be quite significant.

Our results show that some treatments of the Coulomb correction adopted in the
literature are probably inadequate. For example, the CH89 systematics \cite{Va91}
uses the same energy shift $\Delta E_C$ for the real and imaginary OP and it
underestimates, therefore, the Coulomb correction to the imaginary OP as shown
above. The recent global OP by Koning and Delaroche \cite{Kon03} even assumes
$\Delta W_C\approx 0$.

We note that another common usage for Coulomb correction is the entire difference
of the OP for neutrons and protons at the same energy. To show the strength of such a
total Coulomb correction, we generated the 45 MeV proton OP using the same
CH89 formulas but setting $\Delta E_C=0$. This procedure yields
a significantly stronger $\Delta U_C$, with $\Delta V_v/V_v\approx 5.6\%,\
\Delta W_v/W_v\approx 56.9\%,$ and $\Delta W_s/W_s\approx 12.9\%$. However,
one should be careful in discussing such a total Coulomb correction because the CH89
formulas were determined \cite{Va91} using a constant energy shift
$\Delta E_C=19$ MeV for \pPb OP, and it is questionable to use the CH89 potential
obtained with $\Delta E_C=0$ in the present discussion.

\section{Summary}

We have measured cross sections for elastic scattering of neutrons from $^{208}$Pb
at 30.4 and 40.0 MeV and deduced the realistic OP parameters using the Woods-Saxon
geometry given by the CH89 systematics.

The elastic \pPb scattering
data at 45 MeV and elastic \nPb scattering data at 30.4 MeV have about the same
diffractive structure at the forward angles, indicating that the energy shift used
for 45 MeV protons in the CH89 parametrization
should be $\Delta E_C\approx 14.6$ MeV. The isovector part $U_1$ of the nucleon OP
for a $^{208}$Pb target has been used in a CC analysis of the
charge exchange \pn reaction at 45 MeV, exciting the IAS in $^{208}$Bi, and a very
satisfactory description of the \pn data has been obtained. This result allowed
us to use the complex isovector part $U_1$ of the CH89 OP to check the
Lane consistency of the nucleon OP.

The detailed OM analysis of the elastic neutron and proton scattering
has shown that the realistic proton OP at energy $E_p=E_n+\Delta E_C$
contains a non-zero Coulomb correction to its complex strength. Such a non-zero Coulomb
correction represents the isospin impurity of the CH89 nucleon OP, which is only a few
percent for the real part of the OP but is around 20\% for the
imaginary part at the nuclear surface.

We reiterate that these comments are in reference to the CH89 global OP, which already
contains a significant correction for Coulomb effects. These results show that CH89
systematics provides an accurate estimate of Coulomb effects for the real part of the
OP, provided that the energy shifts that enter the model are found by matching
diffraction structures for neutron and proton
scattering. The 20\% correction for the imaginary part of the OP is not large,
and the CH89 systematics can still provide a reasonable priori estimate even for the
imaginary potential.

These results confirm again the importance of elastic neutron scattering
experiments, for use in conjunction with existing elastic proton
scattering and \pn charge exchange  data, to obtain estimates of Coulomb
corrections in heavy nuclei.

\section*{Acknowledgements}
We wish to thank Darrell Drake for the loan of the $^{208}$Pb target. This research was
supported by the U.S. National Science Foundation grants No. PHY-78-22696, PHY06-06007,
PHY08-22648(JINA).  Two of the authors (D.T.K. and B.M.L.) were supported by the National
Foundation for Science and Technology Development (NAFOSTED),
under Project Nr. 103.04-2011.21.

\pagebreak


\begin{thebibliography}{99}
\bibitem{PbT} $^{208}$Pb target was obtained on loan from D. Drake, Los Alamos.
\end{thebibliography}
\end{document}